# Predicting emergence of crystals from amorphous matter with deep learning

Muratahan Aykol,* Amil Merchant, Simon Batzner, Jennifer N. Wei, Ekin Dogus Cubuk*

Google DeepMind

*aykol@google.com, cubuk@google.com

**Crystallization of the amorphous phases into metastable crystals plays a fundamental role in the formation of new matter, from geological to biological processes in nature to synthesis and development of new materials in the laboratory. Predicting the outcome of such phase transitions reliably would enable new research directions in these areas, but has remained beyond reach with molecular modeling or ab-initio methods. Here, we show that crystallization products of amorphous phases can be predicted in any inorganic chemistry by sampling the crystallization pathways of their local structural motifs at the atomistic level using universal deep learning potentials. We show that this approach identifies the crystal structures of polymorphs that initially nucleate from amorphous precursors with high accuracy across a diverse set of material systems, including polymorphic oxides, nitrides, carbides, fluorides, chlorides, chalcogenides, and metal alloys. Our results demonstrate that Ostwald's rule of stages can be exploited mechanistically at the molecular level to predictably access new metastable crystals from the amorphous phase in material synthesis.**

**Main text**

When amorphous solids crystallize, the first phases to emerge are often not the thermodynamic ground states but metastable crystals, in line with Ostwald's rule of stages.[1–5] This rule is ubiquitous in nature. For instance, crystallization of amorphous calcium carbonate to vaterite, aragonite or calcite has significance in biomineralization and carbon cycle.[6,7] Crystallization sequences of amorphous ice or interstellar molecules have importance in geology, astrophysics and astrobiology.[8–10] In materials science, amorphous-to-crystalline transformation underpins the development of new technologies, such as phase change memories,[11] nanocrystallized metallic glass soft magnets,[12] or ceramics with fine-tuned electronic or optical properties.[13] Particularly in synthesis, crystallization of non-crystalline precursors or intermediates plays a central role.[8,14–17] The complexity of nucleation physics has hindered the formulation of polymorph-selective synthesis models,[18] but encouraged the development of numerous experimental techniques to explore the energy landscape such as solid state reaction, chimie douce, deposition, rapid cooling or high pressure methods.[19,20] Unlike polymorphs, an amorphous phase can be accessed selectively with many of these techniques and used as a precursor to metastable crystals.[14,15,21] This instance of Ostwald's rule can be reasoned by classical nucleation theory where crystals which share local structural motifs with the amorphous phase would have a lower barrier against nucleation and appear first upon annealing (Fig. 1a).[1,18,22,23] Given the challenge of controllably navigating the energy landscape in materials synthesis,[15,24,25] this structural connection renders the disordered state a uniquely useful starting point. For the energy landscape itself, state-of-the-art deep learning potentials trained on millions of first-principles calculations are well-poised to provide a reliable description at the scale of atoms.[26–29]

Guided by the omnipresence of amorphous phase and its molecular level structural connection to crystals, here we introduce a computational approach that achieves highly-accurate predictions of metastable crystallization products of amorphous matter in any chemistry. In this approach, we directly model the transformation of local motifs in the disordered states into ordered structures in atomistic simulations using universal deep learning potentials, at a scale impractical for ab-initio methods. Our results show that when combined with modern deep learning, Ostwald's century-old rule unlocks precise *in silico* predictions of the crystal structures emerging from amorphous-to-crystalline phase transitions.

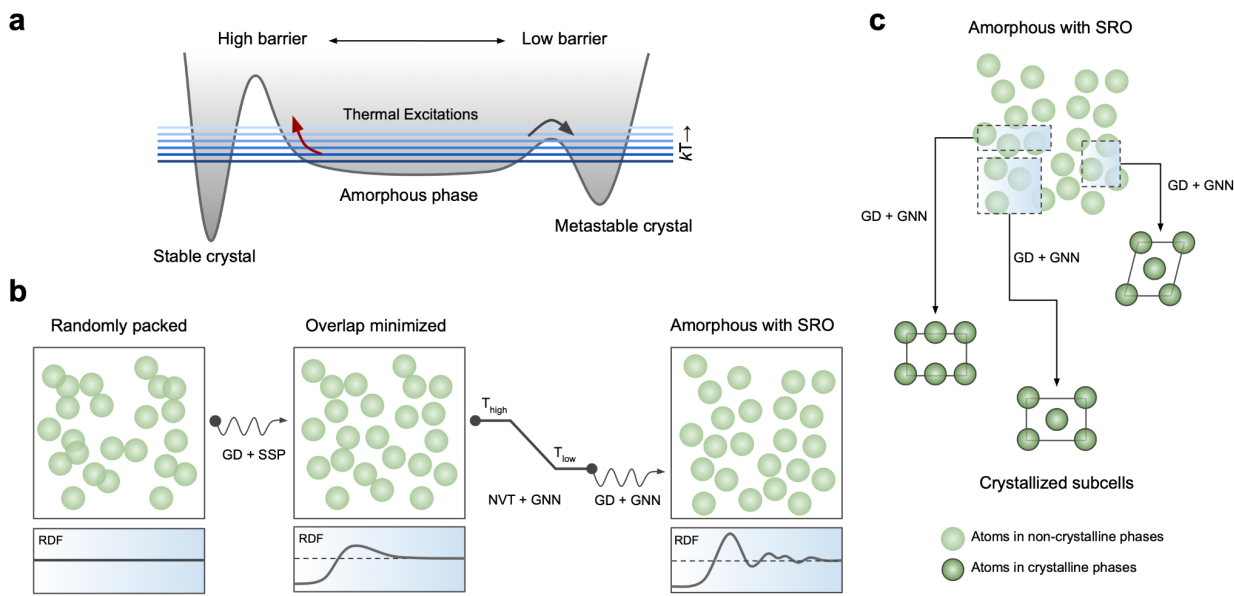

**Figure 1. Predicting crystallization products of the amorphous phase.** (a) Potential energy landscape view of relative ease of accessing the initial and often metastable crystallization product from an amorphous basin compared to a stable crystal, due to the lower barrier to the nucleation of the former crystal. Computational pipeline for (b) generating atomistic models of amorphous phases with properly developed short-range order (SRO) and (c) geometric optimization of subcells of various sizes from the amorphous phase enforcing periodicity. RDF, GD, NVT, SSP and GNN refer to radial distribution function, gradient descent relaxation, constant volume–constant temperature molecular dynamics, soft-sphere interatomic potential, and graph neural network based interatomic potential, respectively.

**Computational framework**

In our approach to predicting the outcome of amorphous-to-crystalline phase transitions (hereafter $a^2c$), we start by creating an atomistic model of the amorphous phase that captures the short-range order via a melt-and-quench molecular dynamics (MQMD)[30] protocol (Fig. 1b). We hypothesize that if the geometric degrees of freedom of a sufficient number of subcells out of the amorphous phase are relaxed under periodic-boundary conditions following downhill gradients on the energy landscape (Fig. 1c), a fraction of those with the seeding motif should relax into the adjacent metastable structure basin that nucleates in practice. The lowest energy configurations reached with this process should be a close match to the crystal structure of the experimentally observed initial crystallization product of the amorphous phase, irrespective of the product being stable or metastable. In our workflow, we typically relax ~$10^4$ to $10^6$ subcells per amorphous system sampled over a grid of parallelepipeds. Performing rapid and accurate relaxations for such large structure pools is enabled by describing interactions with a fast and accurate deep learning interatomic potential, and would be unfeasible to tackle with first principles calculations. Overall, MQMD and crystallization steps in $a^2c$ are designed to balance accuracy with computational speed (Methods).

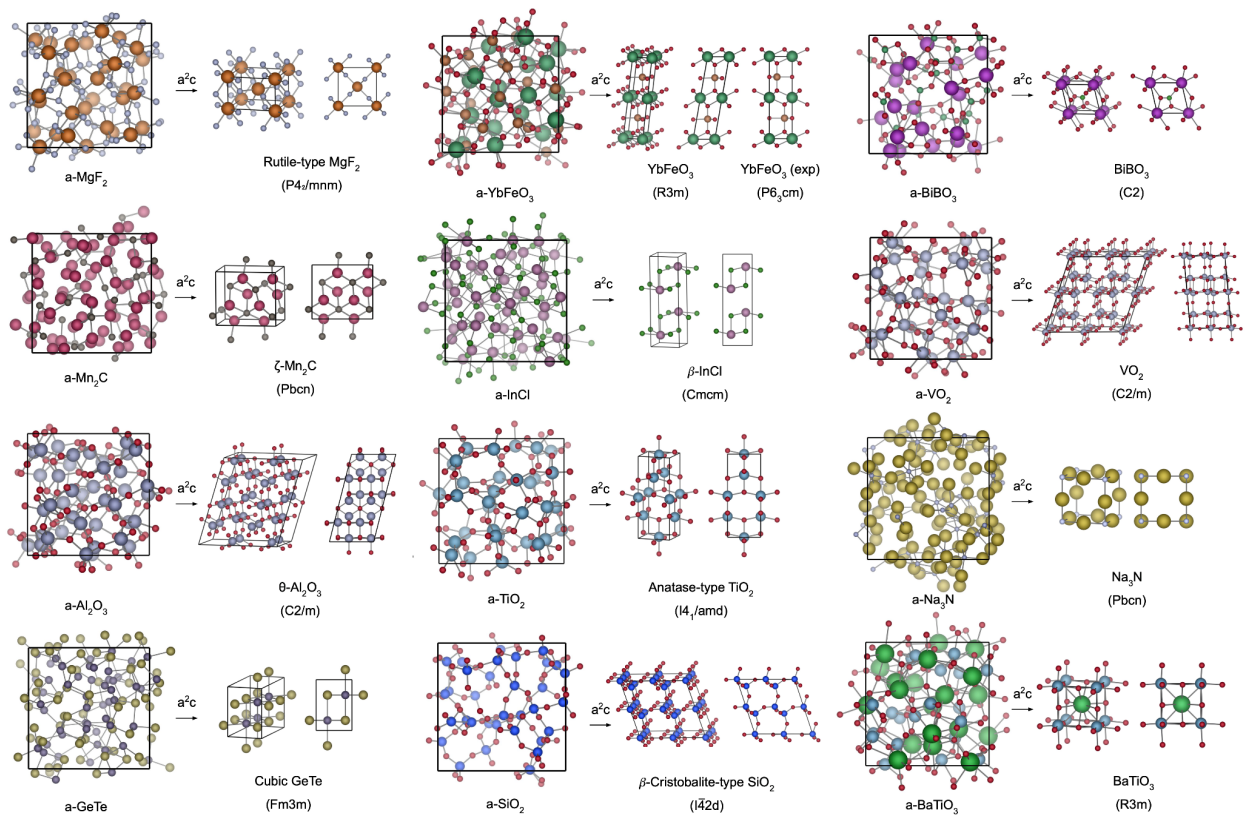

**Figure 2. Crystal structures predicted *de novo* to nucleate from the amorphous phases agree with experiments.** The amorphous phases and the resulting crystal structures found from it using a²c are displayed, where the latter are found to be in remarkable agreement with experimental reports. Crystals are shown at oblique and side view projections. For YbFeO₃ the corresponding experimental structure is also shown in addition to a²c prediction due to the stacking sequence variation. Radial distribution functions corresponding to each of these amorphous and crystal phases are shown in Supplementary Fig. 1.

**Predicting the crystallization products with a²c**

Initial crystallization products predicted using a²c in twelve cases across a range of chemistries are presented in Fig. 2. In each case, a²c sifts through tens of thousands of local motifs in the amorphous phases using deep learning based interactions to identify the shown crystallization products as the lowest energy crystals *accessible from the amorphous phase at the target composition*. These products are metastable in many cases among numerous competing polymorphs, and exist in a practically infinite space of structures, rendering their prediction a needle-in-a-haystack problem.

Metal oxides, given their abundance in nature and as subjects of broad technological interest, offer well-studied polymorphic systems with known crystallization sequences. In the case of amorphous ("a") $TiO_2$, the first crystallization product is anatase, and not rutile, brookite or any other polymorph.[31–33] Before reaching quartz, a-$SiO_2$ visits various polymorphs, the first being cristobalite.[34,35] On crystallization of a-$VO_2$, the less common B-phase appears initially, prior to the ground-state R-phase.[36] Ebralidze et al.[37] found that when the anomalous expansion of a-$BaTiO_3$ films[38] is suppressed, they crystallize into cubic $BaTiO_3$. In all these cases, a²c correctly finds the initial crystallization product of the amorphous phase.

Metal halides, carbides, nitrides and chalcogenides possess different levels of ionic and covalent character, and hence offer other unique polymorphs to test a²c. We find rutile-type $MgF_2$ as the lowest energy amorphous-accessible phase, consistent with annealing of films deposited above -50 °C.[39] Metastable β form of InCl was reported to nucleate on annealing of a-InCl deposited at low temperatures, instead of the stable distorted-rock salt α form,[40] as predicted here. Aouni and Bauer-Grosse[41] discovered the orthorhombic carbide $Mn_2C$ by crystallizing a sputtered amorphous film. They identified the structure to be of ζ-$Fe_2N$ type, in line with the $a^2c$ prediction. For GeTe, $a^2c$ yields the NaCl-type high-temperature β phase, which crystallizes out of a-GeTe instead of the ground state *R3m* phase.[11] Despite the presence of a stable $Li_3N$ phase, the analogous $Na_3N$ phase remained elusive for years.[42] Fischer and Jansen made this highly unstable compound in the anti-$ReO_3$ structure by crystallizing an amorphous precursor deposited on a liquid nitrogen cooled substrate to lock in a highly atomized Na-N mixture.[42] In $a^2c$, the reported deposition temperatures were essential for limiting the development of short-range order and dimerization in the amorphous phase, and in turn for crystallization of anti-$ReO_3$-type $Na_3N$, which is not only validating $a^2c$ but also the ability of tuning the amorphous-phase to induce a target.

When polytypes or disorder are at play, predictions require further discussion. $YbFeO_3$ adopts an orthorhombic perovskite structure in its stable form, but Nishimura et al.[43] synthesized a metastable hexagonal polymorph (*P6₃cm*) from an amorphous precursor, where Fe forms an unusual trigonal bipyramidal coordination with five oxygen atoms, stacked between layers of closed-packed [$YbO_6$]. With $a^2c$, we find a closely-matching polytype in *R3m* that deviates from the original report only in its stacking sequence, hence is nearly isostructural (Fig. 2). Amorphous $Al_2O_3$ goes through topotactically-related transition aluminas on annealing, such as *γ* and θ, before reaching α-corundum. *γ* and θ differ mainly by the degree of site disorder and the ratio of $Al^{3+}$ occupying the tetrahedral and octahedral sites of the cubic-close-packed $O^{2-}$ sublattice. The structure of *γ* phase is still debated,[44] whereas θ is considered a similar but more ordered monoclinic variant. In a-Al$_2$O$_3$ films, while $\gamma$ was historically suggested to crystallize first, recent experiments refined the product to be θ,[45] and showed θ has the closest structural alignment with amorphous alumina among all polymorphs.[46] The prediction from a$^2$c is consistent with these recent reports.

To summarize the implications, a$^2$c is selectively identifying the crystal structure of a crystallizing (and usually metastable) polymorph out of infinitely many stable or metastable possibilities, simply by molecular simulations driven by a deep learning potential, directly from the amorphous phase with no chemical restrictions or explicit input of any particular structure or bias.

**Bridging knowledge gaps with a$^2$c**

Encouraged by the examples in synthesis, we focus next on how a$^2$c can complement experimental observations and draw cases mainly from boron chemistries spanning different bonding characteristics. The first case is solving the structures of crystallizing polymorphs. The bismuth borate system has numerous stable and unstable compounds studied for their optical properties,[47] where two distinct metastable BiBO$_3$ polymorphs labeled (I) and (II) were reported.[48] Polymorph (I) could be accessed by crystallizing the melt and its structure was solved using single-crystal XRD.[49] Polymorph (II), however, could only be synthesized by crystallizing the glass phase and its structure remained unsolved. Recently, Shinozaki et al.[50] analyzed (II) with TEM and XRD and determined it belongs to the monoclinic space group *C2*. With a$^2$c, we find the lowest energy form to be of exactly *C2* symmetry (Fig. 2). The simulated XRD pattern (Supplementary Fig. 2) is indistinguishable from that of Shinozaki et al.,[50] hence completes the structure solution for $BiBO_3$ (II) by including atomic positions.

Next, we use $a^2c$ to understand the mechanism of metastable multiphase crystallization, focusing on the seminal $Fe_{80}B_{20}$ metallic glass.[51,52] Polymorphic (iso-compositional) transitions are common as instances of Ostwald's rule because decomposition into multiple phases with diverging compositions would be penalized by solid-state transport. Under such conditions, we anticipate decomposition of a-$Fe_{80}B_{20}$ to fan out gradually to nearby metastable crystals,[5] consistent with the phase landscape obtained by $a^2c$ (Fig. 3). The first crystal to appear is $Fe_4B$, requiring no long-range transport. Such a transitory phase was reported as super-saturated α-Fe(B) or $Fe_4B$, but its structure has not been established beyond having bcc-like or layered-like features.[53–55] In agreement, we find the $Fe_4B$ phase has distorted bcc-Fe layers separated by layers of tricapped trigonal [$BFe_9$] prisms. Interestingly, this pattern extends to lower Fe contents, forming a homologous series. The well-known metastable crystallization product $Fe_3B$ (*Pnma*)[51,52] is found exactly by $a^2c$, which can be viewed as an extension of the series where [$BFe_9$] layers bond directly. The ground state tetragonal $Fe_2B$ (I4/m) and bcc-Fe phases are also found, which implies their crystallization is bottlenecked by long-range diffusion. These results show that crystallization of a-$Fe_{80}B_{20}$ involves topotactic decomposition to a series of related phases towards Fe and $Fe_3B$, in accord with the spinodal-like phase separation suggested for such glasses.[5,55–57] The $a^2c$ predictions reconcile the available experimental knowledge into a decomposition pathway for $Fe_{80}B_{20}$ and show Ostwald's rule extends to multiphase decomposition when the local structural relationships persist.

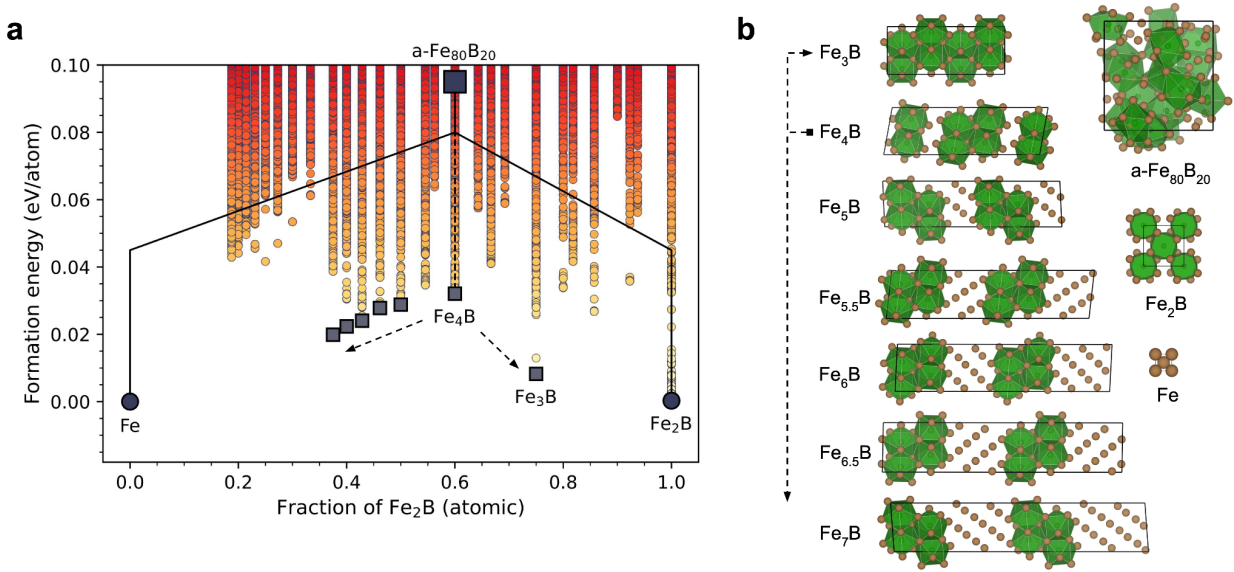

**Figure 3. Crystallization pathway of Fe$_{80}$B$_{20}$ metallic glass as predicted by a$^2$c.** (a) Formation energy of the crystallized phases versus composition, with respect to Fe and Fe$_2$B as terminals. No data outside a$^2$c outputs are used. (b) Atomic structures of the amorphous cell and crystal phases emerged from application of a$^2$c (phases highlighted in (a) with dark blue markers). Green and brown circles represent B and Fe atoms, respectively. In boron-containing structures, B-centered Fe clusters are shown as green polyhedra. In (a), the arrows highlight the decomposition pathway and coloring is a guide for the eye that scales with formation energy. Amorphous Fe$_{80}$B$_{20}$ energy falls outside the shown range and is repositioned for visual purposes.

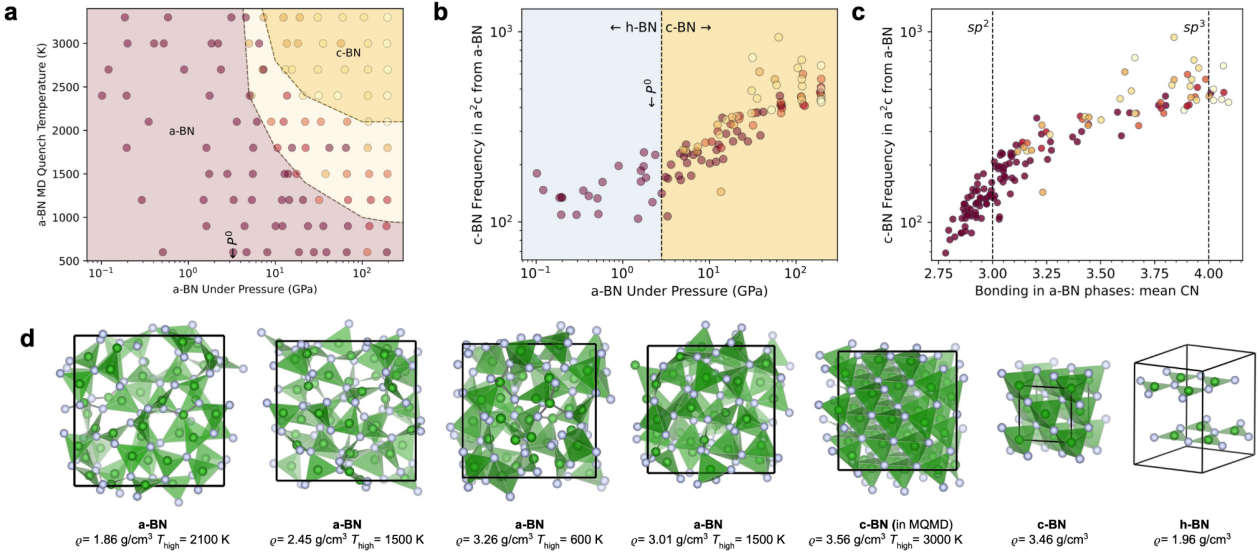

**Figure 4. Polymorph selection through amorphous phase morphology in boron nitride.** (a) Morphology map of structures generated using MQDM as a function of quench temperature and pressure, where circles are colored on a scale ranging from red to yellow corresponding to MQMD runs where the structures in repeating simulations at each parameter combination were 100% amorphous or 100% c-BN, respectively (further details in Methods). The shaded regions outline amorphous, and partially and mostly crystalline regions. (b) Frequency of subcells that converted into c-BN in $a^2c$ for each parent phase shown in (a) as a function of pressure on a-BN or partial or c-BN phases [color coding identical with (a)]. $P^0$ denotes the computationally estimated equilibrium pressure for h ↔ c transition at zero temperature. (c) The same frequency in (b), plotted against the mean B-N coordination number (CN) in the parent MQMD phases. (d) Sample atomic structures of a-BN and c-BN phases that emerged from MQMD, along with the ideal c-BN and h-BN phases. B and N atoms are shown as green and gray spheres, respectively. B-N coordination environments are highlighted with green polyhedra.

The third case is polymorph selection by controlling the morphology of the amorphous phase. Synthesis of $sp^3$-hybridized cubic (c) BN from $sp^2$-hybridized hexagonal (h) BN requires excess pressures and temperatures, whereas starting with a-BN requires less extreme conditions.[58–60] With nominal a-BN phases generated over a fine grid of $P$-$T$ conditions and ~3.1 million subsequent crystallizations, we found h-BN and its nearly-degenerate rhombohedral variants form from the low-density a-BN at lower pressures, and the higher-density c-BN from the higher density a-BN with increasing pressure (Figs. 4a-c). Under extreme pressure and temperature conditions, c-BN appeared even in the MQMD stage (Figs. 4a and 4d). More importantly, we found that the frequency of nominally amorphous BN subcells that crystallize into c-BN from a given a-BN cell increases with pressure (Fig. 4b), which is explainable by the continuous shift from $sp^2$ to $sp^3$ in B-N bonding in a-BN with pressure (Fig. 4c and Supplementary Fig. 3) and provides mechanistic evidence for the easier and direct nucleation of c-BN from a-BN. Another complex polymorph selection case is observed in amorphous calcium carbonate (ACC), where calcite, vaterite and aragonite have all been reported to form first depending on conditions such as $Mg^{2+}$ content, hydration levels or morphology.[7,61,62] Notably, we find all three polymorphs to be accessible from ACCs generated under different conditions (Methods), indicating the sensitivity reported in experiments is present even in the non-aqueous phase-pure ACC structures. These findings, along with $Na_3N$, demonstrate that predictivity in $a^2c$ is retained in polymorph selection under morphological variations in the amorphous parent.

**Elucidating the transformation mechanisms in a$^2$c**

While structural relationships have long been suggested as an empirical explanation for Ostwald's rule, the accuracy achieved with a$^2$c is mechanistic and hence surprising. To understand how local motifs evolve into specific crystals in a$^2$c, we track the progress of a particular subcell in a-$TiO_2$, a model condensed system for Ostwald's step rule,[33] at the atomic-level towards anatase in Fig. 5. The cell starts with suboptimal Ti-Ti and O-O separations and high energy due to periodicity constraint, but the local Ti-O-Ti motif present at ~150° remains as the backbone towards anatase as the cell steadily relaxes to lower energies by local atomic movements and distortions (Fig. 5b to 5e), exclusively on a downhill path. The suboptimal bonds, which would be local excitations activated thermally in annealing, resolve early: first Ti-Ti and then O-O peaks disappear in the first 40 steps (Fig. 4d and 4e) . Beyond this point, structural fingerprints of the cell remain close to that of the amorphous state in both radial and bond angle distribution functions, and Ti centers of the backbone rotate to lock into the anatase arrangement. The peak positions and intensities in distribution functions of the final crystal align with their counterparts in the a-$TiO_2$. In fact, in all systems we studied, the structural fingerprint of the product aligned well with the amorphous parent (Supplementary Fig. 1).

These findings confirm that the local motifs guide the transformation from amorphous phase *geometrically* on a continuous, energetically down-hill path towards a particular polymorph, enabling an unexpected degree of precision at the atomistic level for predicting amorphous-to-crystalline transitions. As seen in Fig. 5, the molecular nature of these transitions is closer to displacive and dilational polymorphic in Buerger's classification,[20,21] which take place between polymorphs with structural relationships under limited mobility (e.g. crystallization at lower temperatures). Unlike the reconstructive cases, these transitions require minimal bond breaking or rearrangement beyond distorations or rotation, and occur via a low-energy interface due to structural similarity of the involved polymorphs. Therefore, these transitions proceed rapidly,[20] where any nucleus forming cascades into its neighborhood in the amorphous matrix. This lean mechanistic nature coincides with what is modeled atomistically with $a^2c$, hence explains why its predictions are consistent with experimental observations in many different chemistries.

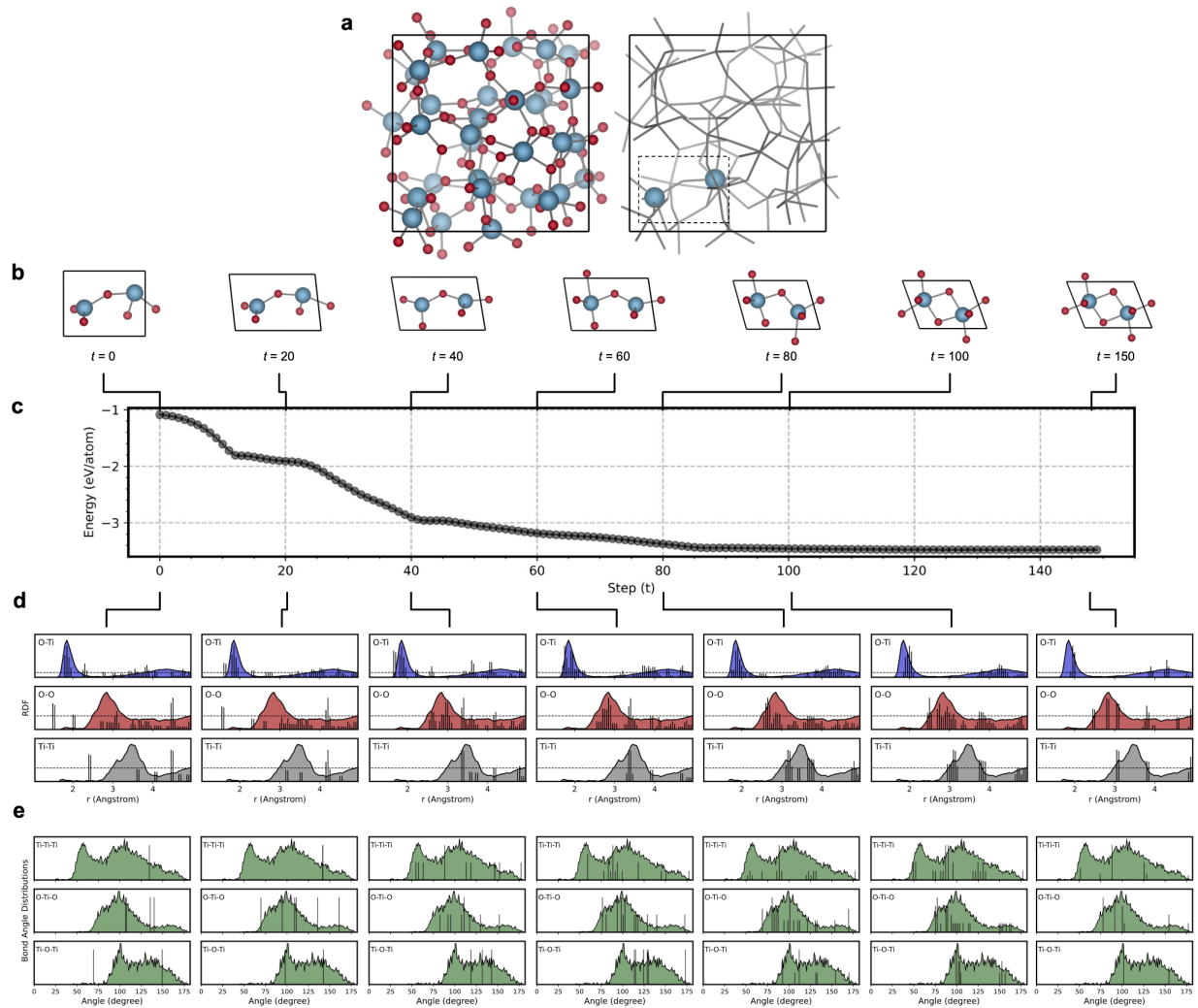

**Figure 5. Emergence of anatase from a local motif in amorphous TiO$_2$.** (a) The amorphous structure for TiO$_2$ obtained via MQMD, the first panel showing the entire atomic structure and the second one highlighting the subcell (and its two Ti atoms) being tracked. Evolution of the geometry (b) and energy (c) of the subcell in (a) upon optimization using the GNN-potential under the periodic-boundary constraint leading to the primitive anatase cell. The radial and bond angle distribution functions of the subcells in (b) are shown in (d) and (e) as vertical solid lines, respectively, both overlaying the same functions of the parent amorphous phase. Intensities are scaled to the same ranges for amorphous and crystal functions for visualization. For consistency across amorphous and crystal phases, Ti-O, O-O and Ti-Ti maximum bond lengths were approximated as the first minima of the RDF as 3.5, 2.5 and 4.0 Å, respectively.

**Conclusion and outlook**

Our work has demonstrated that zero-shot prediction of the products of amorphous-to-crystalline phase transitions *de novo* in any chemistry with molecular-level explainability is possible by combining local motif search with the new general-purpose deep learning interatomic potentials. Predictive agreement with experimental polymorphs across a wide range of chemistries validated both the computational framework, and the structural argument commonly made in support of Ostwald's rule of stages for amorphous-to-crystalline transformations. Provided this validation, the presented approach can now be employed in many new research directions, such as to resolve the nature of such transitions in complex material systems, or towards *in silico* discovery of material polymorphs or glass-crystal composite systems that inherently have concrete synthetic pathways from the amorphous state.

**Methods**

*Generation of amorphous structures*

The melt and quench molecular dynamics (MQMD) process in a$^2$c highlighted in Fig. 1b aims to obtain reasonably representative amorphous configurations with an efficient procedure in three stages. In the first stage, *N* atoms comprising the system are distributed randomly in a cubic box. Except when explicitly modeling variations in density, the density of this box is determined heuristically with the rule-of-thumb of 15% volumetric expansion of the volume/per atom of the known stable phase or averaging the same in case of the presence of a phase mixture at that composition on the convex hull in our standard workflow. This rule-of-thumb is similar to those previously reported in the literature,[30] and yields consistent amorphous configurations with no cavities. For maintaining throughput and efficiency, N is restricted to the exact integral representation of the composition closest to 100, which was shown to yield sufficiently representative structures in inorganic systems.[30] The random atomic positions are then relaxed briefly using an optimization algorithm called FIRE[63] (which is used for all GD optimizations in this work) and a soft-sphere interaction potential to minimize the atomic overlaps. This stage ensures the subsequent amorphous structures do not inherit fingerprints from any particular crystal and instead rapidly develop the proper short-range inherent to liquid and amorphous phases in the equilibration and quench stages.[30] The number of ionic steps in overlap minimization is limited to 30, or stopped early when the shortest interatomic distance reaches above the 90% of summed ionic radii of the atoms of each pair. Keeping the number of steps limited in this process ensures that we do not inject any arbitrary local arrangement using the soft-sphere potential. Next stage in the workflow is a three step MD protocol where a GNN-based force field is used throughout: equilibrate a melt at $T_{high}$, cool/quench this phase to $T_{low}$, and equilibrate at this temperature. $T_{high}$ and $T_{low}$ were manually selected for each system (Supplementary Table 1), and we ran each of the three stages of the protocol for 1500, 2500 and 1500 steps, respectively. In MD, we use the *NVT* ensemble with a Nose-Hoover thermostat and a timestep of 2 fs. The atomic positions can be relaxed with a short GD run in the last stage, which was kept optional as we did not find any impact on crystallization. The MQMD workflow is implemented in JAX-MD,[64] and can be run in high-throughput mode.[65]

*Crystallization workflow*

The crystallization phase in a²c (Fig. 1c) accepts the amorphous structure model generated in the MQMD as input, and exhaustively creates all orthorhombic parallelepipeds (subcells) that can be generated from this structure, keeping their edges parallel to the unit vectors of the parent cubic box. These subcells are allowed to be of any length and at any position on the $n_{grid} \times n_{grid} \times n_{grid}$ point net [first point located at the position (0, 0, 0)] encompassing the parent structure. Our default for $n_{grid}$ is 10, except $Fe_{80}B_{20}$ system and $CaCO_3$ as noted in Supplementary Table 1. Additional degrees of freedom can be added to the subcell creation, for instance by allowing triclinic cells or shifts, but we found the present setting to yield a comprehensive-enough set of subcells to search for crystal seeds: the full grid has ~$1.7 \times 10^5$ subcells for $n_{grid}$ = 10 and ~$1.0 \times 10^7$ subcells for $n_{grid}$ = 20. These subcells can be filtered further using maximum atom counts per cell or stoichiometry constraints in cases where compositions distant from the parent amorphous phase are not of interest or not plausible. For example, in $Fe_{80}B_{20}$, all subcells are kept up to an atom count of 16, whereas for $MgF_2$, subcells having this target stoichiometry and up to 21 atoms are kept. With the stoichiometry and atom-count constraints, subcell set size per system often falls in the range $10^3 - 10^4$ for most systems with $n_{grid}$ = 10 and 1-2 orders of magnitude larger for $n_{grid}$ = 20. Details on each system studied are provided in Supplementary Table 1. Next, all geometric degrees of freedom of these subcells, including atomic positions and the cell size and shape, are optimized using a NequIP based equivariant GNN-potential (See Deep learning methods) to describe interatomic interactions, and the previously mentioned

GD algorithm[63] for at least 150 steps. The lowest energy structures found by this procedure at each composition are the *selectively-predicted* stable or metastable crystallization products that should match the experiments. Our lowest-energy bound is strict, i.e. experimental crystallization products were either the absolute lowest energy a²c predictions, or practically degenerate within 1-2 meV/atom, which is on par or lower than the variability in DFT convergence and/or structure optimization, hence negligible. The crystallization process yields an extremely large pool of structures spanning motifs not only relevant for the target compound, but also nearby compounds, as exemplified in Supplementary Fig. 4 for cases where stoichiometry constraint was not applied in a²c to observe the extend of compositional and energetic diversity accessible with the process.

*Deep learning methods*

General-purpose machine learning interatomic potentials of various deep learning architectures are emerging for modeling condensed phases,[26–29,66,67,70] certain architectures being chemically universal (i.e. a single potential modeling chemistries spanning the entire periodic table and under various conditions). The particular instance of the general-purpose deep learning potential used in this work adopts the equivariant GNN architecture of NequIP,[68] and uses 3 layers of message passing, even irreps up to $l_{max}$ of 2 with multiplicities 64, 32, 16 for $l$=0, 1, 2, a two-hidden-layer radial multilayer perceptron with 64 neurons acting on a radial basis of eight Bessel functions. We use a local interatomic cutoff radius of 5 Å in graph generation. This particular model has ~2.4 million parameters trained on energies and forces with a loss function composed of equally-weighted Huber losses where the energy-term is normalized per-atom and the force term is normalized by the total number of force components 3N, each with a Huber-loss delta of 0.01. The potential was trained using the Adam optimizer with a batch size of 64 structures and a learning of $5 \times 10^{-3}$ using a linear learning rate decay. We trained our potential on a dataset of several million DFT calculations across a wide array of chemistries and including both ab-initio random search generated structures,[69] and substitutionally-generated structures.[28] We further showed this potential is robust in modeling amorphous systems, attaining a remarkable accuracy of 38 meV/atom testing error with respect to DFT for thousands of amorphous structures encompassing nearly all practical binary chemical systems.[65] This same potential was used in all results reported in this work, including both MQMD and crystallization workflows. All MQMD and crystallization simulations were run on a single P100 GPU.

*Boron nitride search*

To identify the effect of quench temperature ($T_{high}$) and applied pressure ($P$) on the morphology of the *targeted* amorphous BN phase generation, and the resulting crystallization process, we ran MQMD workflow over a grid of 15 densities ($\varrho$) between 1.70 and 4.35 g/cm$^3$ and 10 $T_{high}$ values between 600 and 3300 K, with 5 repeats with different seeds for each each ($P, T_{high}$) combination. In total, 750 MQMD simulations were run for BN. Resulting structures were inspected manually to check if structure is amorphized, or if there are occurrences of partial or full crystallization. Pressure for each point is calculated by averaging that of its 5 repeats. Coordination numbers were determined using Brunner's reciprocal gap method.[71,72] For each ($P, T_{high}$) combination, one of the resulting structures was passed on to the crystallization workflow (Fig. 1). Each of these 150 crystallization runs processed around 20,500 subcells embedded in their parent structure, hence the total number for BN reached ~3.1 million structure optimizations using the GNN potential. For the reference 0 K equilibrium pressure for $h \leftrightarrow c$ transition, which is still under debate,[73] we used the energy difference between $h$ and $c$ phases as calculated with density functional theory[74] and a simple linear approximation to the PV term in enthalpy,[75] which agrees reasonably with the latest diffusion Monte Carlo calculations[73] and is shown as a guide for the eye. We refer to all disordered BN structures as a-BN, including those that display a turbostratic $sp^2$ nature (e.g. $\varrho$ = 1.86 g/cm$^3$ and $T_{high}$ = 2100 K in Fig. 4c).

*Calcium carbonate search*

To obtain idealized, pure-phase ACC structures with varying morphologies and degrees of local order, we ran MQMD over a 3 x 3 grid of densities ($\varrho$ = 2.09, 2.27 and 2.61 g/cm$^3$) and melt equilibration temperatures ($T_{high}$ = 500, 750 and 1150 K). Each of these amorphous structures are subsequently subjected to crystallization (Supplementary Table 1). Calcite appeared from ACC under all conditions, aragonite appeared in mid and high density samples, and vaterite appeared in high density, mid temperature sample. The significance of this result is that each of these polymorphs are accessible from even the idealized, non-aqueous ACC depending on its history (or morphology). While calcite and aragonite structures are well-known, vaterite's structure is not established and has multiple proposals.[76–79] We therefore included several such model structures proposed for vaterite in structure matching in analyzing the crystallization results.

*Structural analysis and post-processing*

Pymatgen is used extensively to process and analyze amorphous and crystal structures.[80] VESTA is used to generate the structure visuals.[81] For smoothness, RDFs and bond angle distributions for amorphous structures are averaged over the last 200 steps of the low temperature equilibration run, in 10 step increments.

**References**

1. De Yoreo, J. J. Casting a bright light on Ostwald's rule of stages. *Proc. Natl. Acad. Sci. U. S. A.* **119**, (2022).

2. Ostwald, W. Studien über die Bildung und Umwandlung fester Körper: 1. Abhandlung: Übersättigung und Überkaltung. *Zeitschrift für Physikalische Chemie* **22U**, 289–330 (1897).

3. Chung, S.-Y., Kim, Y.-M., Kim, J.-G. & Kim, Y.-J. Multiphase transformation and Ostwald's rule of stages during crystallization of a metal phosphate. *Nat. Phys.* **5**, 68–73 (2008).

4. De Yoreo, J. J. Principles of crystal nucleation and growth. *Rev. Mineral. Geochem.* **54**, 57–93 (2003).

5. Schmelzer, J. W. P. & Abyzov, A. S. How Do Crystals Nucleate and Grow: Ostwald's Rule of Stages and Beyond. in *Thermal Physics and Thermal Analysis* (eds. Hubík, Š. J. & Mareš J., P.) 195–211 (Springer, Cham, 2017).

6. Rodriguez-Blanco, J. D., Shaw, S. & Benning, L. G. The kinetics and mechanisms of amorphous calcium carbonate (ACC) crystallization to calcite, via vaterite. *Nanoscale* **3**, 265–271 (2011).

7. Radha, A. V., Forbes, T. Z., Killian, C. E., Gilbert, P. U. P. A. & Navrotsky, A. Transformation and crystallization energetics of synthetic and biogenic amorphous calcium carbonate. *Proc. Natl. Acad. Sci. U. S. A.* **107**, 16438–16443 (2010).

8. Janicki, T. D., Wan, Z., Liu, R., Evans, P. G. & Schmidt, J. R. Guiding epitaxial crystallization of amorphous solids at the nanoscale: Interfaces, stress, and precrystalline order. *J. Chem. Phys.* **157**, 100901 (2022).

9. Jenniskens, P. & Blake, D. F. Crystallization of amorphous water ice in the solar system. *The Astrophysical Journal* **473**, 1104–1113 (1996).

10. Hudson, R. L. Infrared spectra and band strengths of CH3SH, an interstellar molecule. *Phys. Chem. Chem. Phys.* **18**, 25756–25763 (2016).

11. Nonaka, T., Ohbayashi, G., Toriumi, Y., Mori, Y. & Hashimoto, H. Crystal structure of GeTe and Ge2Sb2Te5 meta-stable phase. *Thin Solid Films* **370**, 258–261 (2000).

12. McHenry, M. E., Willard, M. A. & Laughlin, D. E. Amorphous and nanocrystalline materials for applications as soft magnets. *Prog. Mater Sci.* **44**, 291–433 (1999).

13. Holand, W. & Beall, G. H. *Glass-Ceramic Technology*. (John Wiley & Sons, 2019).

14. Johnson, D. C. Controlled synthesis of new compounds using modulated elemental reactants. *Curr. Opin. Solid State Mater. Sci.* **3**, 159–167 (1998).

15. Cordova, D. L. M. & Johnson, D. C. Synthesis of Metastable Inorganic Solids with Extended Structures. *Chemphyschem* **21**, 1345–1368 (2020).

16. Evans, P. G., Chen, Y., Tilka, J. A., Babcock, S. E. & Kuech, T. F. Crystallization of

amorphous complex oxides: New geometries and new compositions via solid phase epitaxy. *Curr. Opin. Solid State Mater. Sci.* **22**, 229–242 (2018).

17. Addadi, L., Raz, S. & Weiner, S. Taking advantage of disorder: Amorphous calcium carbonate and its roles in biomineralization. *Adv. Mater.* **15**, 959–970 (2003).

18. Aykol, M., Montoya, J. H. & Hummelshøj, J. Rational Solid-State Synthesis Routes for Inorganic Materials. *J. Am. Chem. Soc.* **143**, 9244–9259 (2021).

19. Kohlmann, H. Looking into the black box of solid‐state synthesis. *Eur. J. Inorg. Chem.* **2019**, 4174–4180 (2019).

20. West, A. R. *Solid State Chemistry and its Applications*. (John Wiley & Sons, 2022).

21. Stoch, L. & Waclawska, I. Phase Transformations in Amorphous Solids. *High Temp. Mater. Processes* **13**, 181–202 (1994).

22. Threlfall, T. Structural and Thermodynamic Explanations of Ostwald's Rule. *Org. Process Res. Dev.* **7**, 1017–1027 (2003).

23. Bauers, S. R. *et al.* Structural Evolution of Iron Antimonides from Amorphous Precursors to Crystalline Products Studied by Total Scattering Techniques. *J. Am. Chem. Soc.* **137**, 9652–9658 (2015).

24. Schön, J. C. & Jansen, M. Determination, prediction, and understanding of structures, using the energy landscapes of chemical systems - Part I. *Zeitschrift für Kristallographie - Crystalline Materials* **216**, 307–325 (2001).

25. Jansen, M. A concept for synthesis planning in solid-state chemistry. *Angew. Chem. Int. Ed Engl.* **41**, 3746–3766 (2002).

26. Deringer, V. L. *et al.* Origins of structural and electronic transitions in disordered silicon. *Nature* **589**, 59–64 (2021).

27. Chen, C. & Ong, S. P. A universal graph deep learning interatomic potential for the periodic table. *Nature Computational Science* **2**, 718–728 (2022).

28. Merchant, A. *et al. Submitted* (2023).

29. Deringer, V. L., Caro, M. A. & Csányi, G. Machine Learning Interatomic Potentials as Emerging Tools for Materials Science. *Adv. Mater.* **31**, e1902765 (2019).

30. Aykol, M., Dwaraknath, S. S., Sun, W. & Persson, K. A. Thermodynamic limit for synthesis of metastable inorganic materials. *Sci Adv* **4**, eaaq0148 (2018).

31. Lin, C.-P., Chen, H., Nakaruk, A., Koshy, P. & Sorrell, C. C. Effect of Annealing Temperature on the Photocatalytic Activity of TiO2 Thin Films. *Energy Procedia* **34**, 627–636 (2013).

32. Abbasi, M. *et al.* In situ observation of medium range ordering and crystallization of amorphous TiO2 ultrathin films grown by atomic layer deposition. *APL Mater.* **11**, 011102 (2023).

33. Matthews, A. The crystallization of anatase and rutile from amorphous titanium dioxide under hydrothermal conditions. *American Mineralogist* **61**, 419–424 (1976).

34. Bettermann, P. & Liebau, F. The transformation of amorphous silica to crystalline silica under hydrothermal conditions. *Contrib. Mineral. Petrol.* **53**, 25–36 (1975).

35. Carr, R. M. & Fyfe, W. S. Some observations on the crystallization of amorphous silica. *American Mineralogist: Journal of Earth and Planetary Materials* **43**, 908–916 (1958).

36. Stone, K. H. *et al.* Influence of amorphous structure on polymorphism in vanadia. *APL Mater.* **4**, 076103 (2016).

37. Ebralidze, I., Lyahovitskaya, V., Zon, I., Wachtel, E. & Lubomirsky, I. Anomalous

pre-nucleation volume expansion of amorphous BaTiO3. *J. Mater. Chem.* **15**, 4258–4261 (2005).

38. Lyahovitskaya, V. *et al.* Formation and thermal stability of quasi-amorphous thin films. *Phys. Rev. B Condens. Matter* **71**, 094205 (2005).

39. Bach, A. *et al.* Structural Evolution of Magnesium Difluoride: from an Amorphous Deposit to a New Polymorph. *Inorg. Chem.* **50**, 1563–1569 (2011).

40. Bach, A., Fischer, D. & Jansen, M. Metastable Phase Formation of Indium Monochloride from an Amorphous Feedstock. *Z. Anorg. Allg. Chem.* **639**, 465–467 (2013).

41. Bauer-Grosse, E. & Aouni, A. New ζ-Mn2C hemicarbide formed during the crystallization of an amorphous sputtered Mn1−xCx film. *J. Alloys Compd.* **336**, 190–195 (2002).

42. Fischer, D. & Jansen, M. Synthesis and Structure of Na3N. *Angew. Chem. Int. Ed.* **41**, 1755–1756 (2002).

43. Nishimura, T., Hosokawa, S., Masuda, Y., Wada, K. & Inoue, M. Synthesis of metastable rare-earth–iron mixed oxide with the hexagonal crystal structure. *J. Solid State Chem.* **197**, 402–407 (2013).

44. Ayoola, H. O. *et al.* Evaluating the accuracy of common γ-Al2O3 structure models by selected area electron diffraction from high-quality crystalline γ-Al2O3. *Acta Mater.* **182**, 257–266 (2020).

45. Broas, M., Kanninen, O., Vuorinen, V., Tilli, M. & Paulasto-Kröckel, M. Chemically Stable Atomic-Layer-Deposited Al2O3 Films for Processability. *ACS Omega* **2**, 3390–3398 (2017).

46. Pugliese, A. *et al.* Atomic-Layer-Deposited Aluminum Oxide Thin Films Probed with X-ray Scattering and Compared to Molecular Dynamics and Density Functional Theory Models. *ACS Omega* **7**, 41033–41043 (2022).

47. Ihara, R., Honma, T., Benino, Y., Fujiwara, T. & Komatsu, T. Second-order optical nonlinearities of metastable BiBO3 phases in crystallized glasses. *Opt. Mater.* **27**, 403–408 (2004).

48. Pottier, M. J. Mise en Évidence D'Un Composé BiBo3 et de Son Polymorphisme par Spectroscopie Vibrationnelle. *Bull. Soc. Chim. Belg.* **83**, 235–238 (1973).

49. Becker, P. & Fröhlich, R. Crystal growth and crystal structure of the metastable bismuth orthoborate BiBO3. *Z. Naturforsch. B: J. Chem. Sci.* **59**, 256–258 (2004).

50. Shinozaki, K., Hashimoto, K., Honma, T. & Komatsu, T. TEM analysis for crystal structure of metastable BiBO3 (II) phase formed in glass by laser-induced crystallization. *J. Eur. Ceram. Soc.* **35**, 2541–2546 (2015).

51. Duhaj, P., Švec, P. & Zemčík, T. Micromechanism of crystallization of Fe80B20 amorphous alloy. *Mater. Lett.* **9**, 235–241 (1990).

52. Zhang, Y. D. *et al.* Crystallization of Fe‐B amorphous alloys: A NMR and x‐ray study. *J. Appl. Phys.* **61**, 3231–3233 (1987).

53. Duhaj, P. & Svec, P. Formation of metastable phases from amorphous state. *Materials Science and Engineering A* **226228**, 245–254 (1997).

54. Köster, U. & Herold, U. Crystallization of amorphous Fe80B20. *Scr. Metall.* **12**, 75–77 (1978).

55. Aykol, M., Mekhrabov, A. O. & Akdeniz, M. V. Nano-scale phase separation in amorphous Fe–B alloys: Atomic and cluster ordering. *Acta Mater.* **57**, 171–181

(2009).

56. Hirotsu, Y. *et al.* Nanoscale phase separation in metallic glasses studied by advanced electron microscopy techniques. *Intermetallics* **12**, 1081–1088 (2004).

57. Yao, K. F. & Zhang, C. Q. Fe-based bulk metallic glass with high plasticity. *Appl. Phys. Lett.* **90**, 061901 (2007).

58. Huang, J. Y. & Zhu, Y. T. Atomic-Scale Structural Investigations on the Nucleation of Cubic Boron Nitride from Amorphous Boron Nitride under High Pressures and Temperatures. *Chem. Mater.* **14**, 1873–1878 (2002).

59. Gladkaya, I. S., Kremkova, G. N. & Slesarev, V. N. Turbostratic boron nitride (BNt) under high pressures and temperatures. *Journal of the Less Common Metals* **117**, 241–245 (1986).

60. Sumiya, H., Iseki, T. & Onodera, A. High pressure synthesis of cubic boron nitride from amorphous state. *Mater. Res. Bull.* **18**, 1203–1207 (1983).

61. Nielsen, M. H., Aloni, S. & De Yoreo, J. J. In situ TEM imaging of $CaCO_3$ nucleation reveals coexistence of direct and indirect pathways. *Science* **345**, 1158–1162 (2014).

62. Sun, W., Jayaraman, S., Chen, W., Persson, K. A. & Ceder, G. Nucleation of metastable aragonite $CaCO_3$ in seawater. *Proc. Natl. Acad. Sci. U. S. A.* **112**, 3199–3204 (2015).

63. Bitzek, E., Koskinen, P., Gähler, F., Moseler, M. & Gumbsch, P. Structural relaxation made simple. *Phys. Rev. Lett.* **97**, 170201 (2006).

64. Schoenholz, S. S. & Cubuk, E. D. JAX, M.D.: A Framework for Differentiable Physics. *arXiv [physics.comp-ph]* (2019).

65. Aykol, M., Wei, J. N., Batzner, S., Merchant, A. & Cubuk, E. D. Predicting Properties of Amorphous Solids with Graph Network Potentials. in *1st Workshop on the Synergy of Scientific and Machine Learning Modeling @ ICML2023* (2023).

66. Artrith, N. & Urban, A. An implementation of artificial neural-network potentials for atomistic materials simulations: Performance for TiO2. *Comput. Mater. Sci.* **114**, 135–150 (2016).

67. Deng, B. *et al.* CHGNet as a pretrained universal neural network potential for charge-informed atomistic modelling. *Nature Machine Intelligence* 1–11 (2023).

68. Batzner, S. *et al.* E(3)-equivariant graph neural networks for data-efficient and accurate interatomic potentials. *Nat. Commun.* **13**, 2453 (2022).

69. Pickard, C. J. & Needs, R. J. Ab initio random structure searching. *J. Phys. Condens. Matter* **23**, 053201 (2011).

70. Riebesell, J. *et al.* Matbench Discovery -- An evaluation framework for machine learning crystal stability prediction. *arXiv [cond-mat.mtrl-sci]* (2023).

71. Brunner, G. O. A definition of coordination and its relevance in the structure types AlB2 and NiAs. *Acta Crystallogr. A* **33**, 226–227 (1977).

72. Pan, H. *et al.* Benchmarking Coordination Number Prediction Algorithms on Inorganic Crystal Structures. *Inorg. Chem.* **60**, 1590–1603 (2021).

73. Nikaido, Y. *et al.* Diffusion Monte Carlo Study on Relative Stabilities of Boron Nitride Polymorphs. *J. Phys. Chem. C* **126**, 6000–6007 (2022).

74. Jain, A. *et al.* Commentary: The Materials Project: A materials genome approach to accelerating materials innovation. *APL Mater.* **1**, 011002 (2013).

75. Amsler, M., Hegde, V. I., Jacobsen, S. D. & Wolverton, C. Exploring the

High-Pressure Materials Genome. *Phys. Rev. X* **8**, 041021 (2018).

76. Meyer, H. J. Struktur und Fehlordnung des Vaterits. *Zeitschrift für Kristallographie - Crystalline Materials* **128**, 183–212 (1969).

77. Demichelis, R., Raiteri, P., Gale, J. D. & Dovesi, R. A new structural model for disorder in vaterite from first-principles calculations. *CrystEngComm* **14**, 44–47 (2011).

78. Christy, A. G. A Review of the Structures of Vaterite: The Impossible, the Possible, and the Likely. *Cryst. Growth Des.* **17**, 3567–3578 (2017).

79. Medeiros, S. K., Albuquerque, E. L., Maia, F. F., Caetano, E. W. S. & Freire, V. N. First-principles calculations of structural, electronic, and optical absorption properties of $CaCO_3$ Vaterite. *Chem. Phys. Lett.* **435**, 59–64 (2007).

80. Ong, S. P. *et al.* Python Materials Genomics (pymatgen): A robust, open-source python library for materials analysis. *Comput. Mater. Sci.* **68**, 314–319 (2013).

81. Momma, K. & Izumi, F. VESTA 3 for three-dimensional visualization of crystal, volumetric and morphology data. *J. Appl. Crystallogr.* **44**, 1272–1276 (2011).

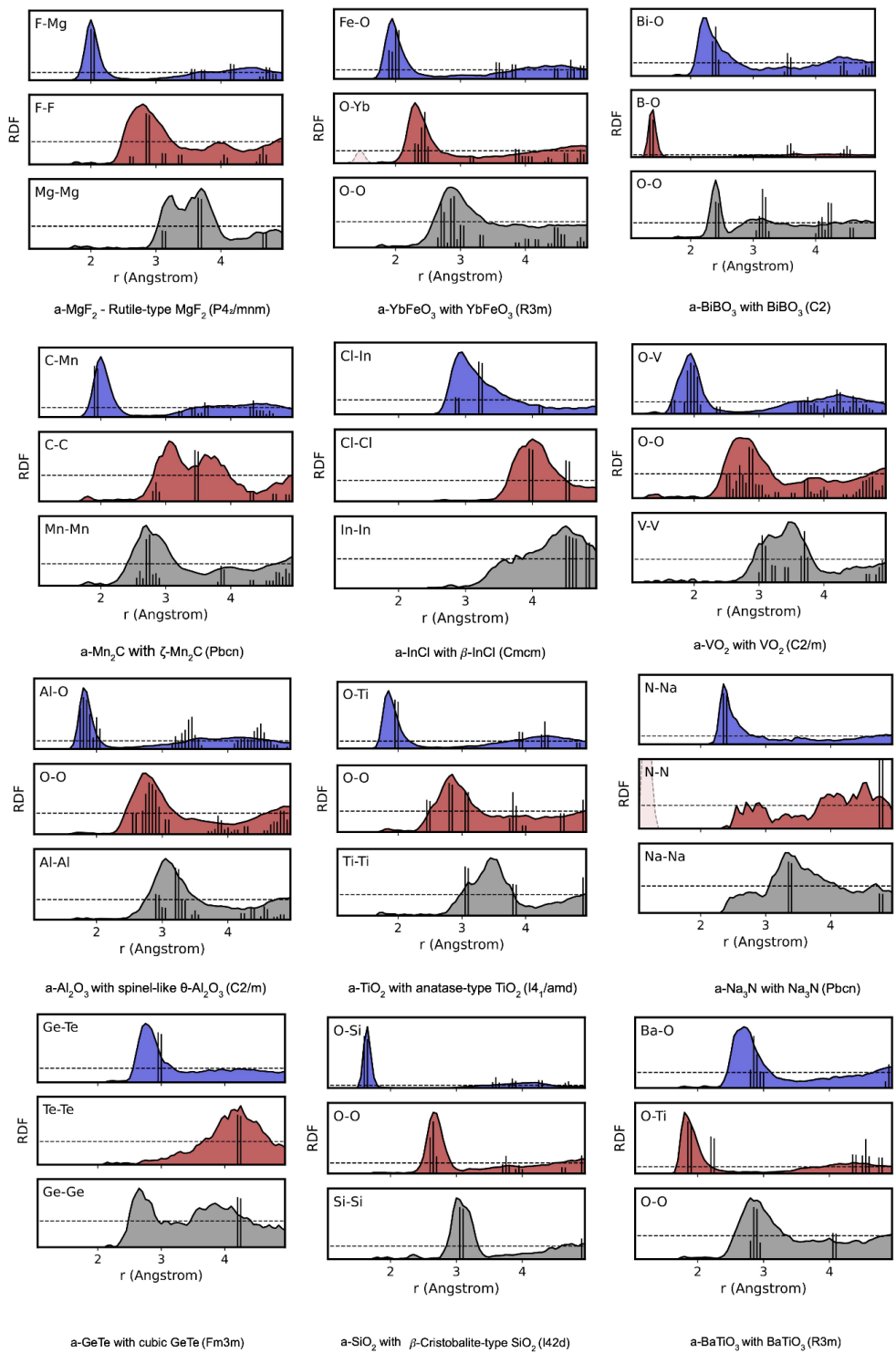

**Supplementary Figure 1. Local order in amorphous and crystalline phases.** Radial distribution functions (RDFs) are plotted for each amorphous phase shown for the systems in Fig. 2, along with the same function of their initial crystallization products.

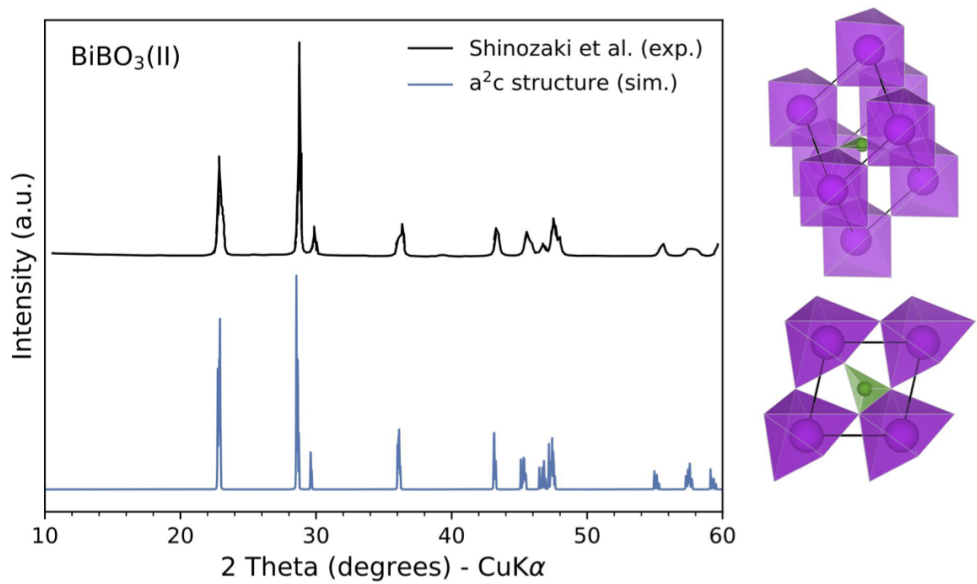

**Supplementary Figure 2. XRD pattern of a$^2$c predicted *C2* BiBO$_3$ structure compared to experiment.** The experimental XRD pattern was extracted from Shinozaki et al.[50] manually by image processing. Polyhedral views of the crystal structure are also shown as inset, where purple polyhedra are [BiO$_6$] clusters, green triangle is the [BO$_3$] units and O-atoms are not displayed for clarity.

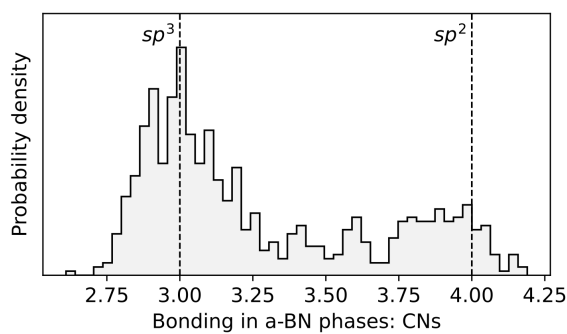

**Supplementary Figure 3. Coordination numbers in a-BN.** Distribution is calculated from the 750 BN phases obtained with the MQMD process as described in Methods.

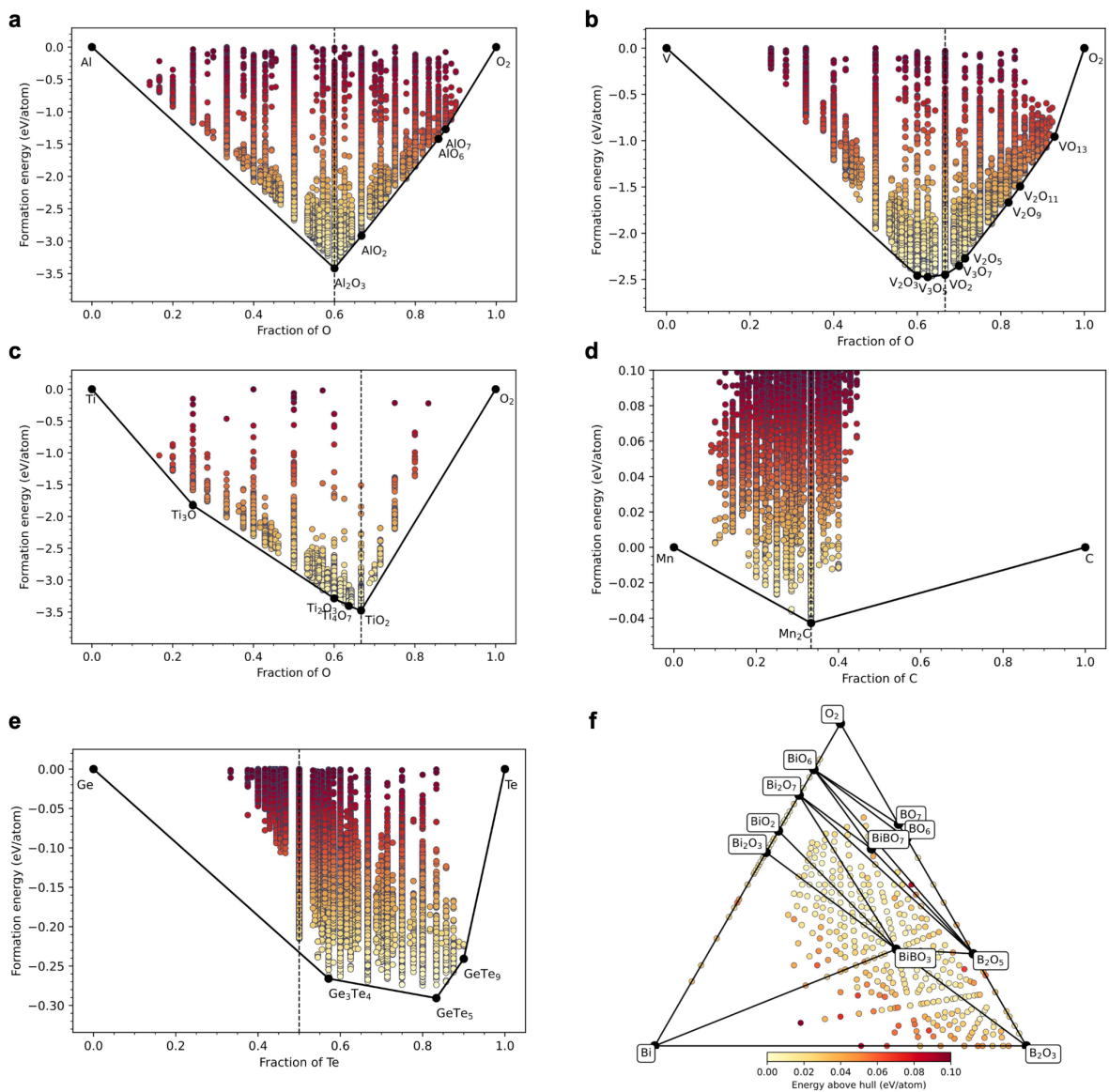

**Supplementary Figure 4. Phase spaces generated from amorphous precursors.** Composition versus energy diagrams are shown where no stable or metastable phase other than those generated in a$^2$c are included. Diagrams (a) to (e) show binary spaces, where composition of the amorphous parent is indicated by a dashed line. Coloring is a guide to the eye that scales with the formation energy. Diagram (f) shows binary and ternary phases generated from a-$BiBO_3$, where each circle represents the lowest energy phase at that composition and color by its distance to the metastable hull generated by a$^2$c phases. The convex hulls serve as a guide only and do not imply thermodynamic ground state and/or immediate decomposition of the amorphous phase.

**Supplementary Table 1.** Settings and statistics for the studied systems. If no stoichiometry constraint was applied, the counts for both the relaxed subcells with target composition (i.e. the same composition as the amorphous phase) and all relaxed subcells are shown in the last column.

| System | $T_{high}$ (K) | $T_{low}$ (K) | Volume (Å$^3$/atom) | $n_{max}$ | Stoichiometry constraint | Subcells relaxed with GNN |
|---|---|---|---|---|---|---|
| $MgF_2$ | 900 | 150 | 12.50 | 21 | Yes | $1.5 \times 10^4$ |
| $YbFeO_3$ | 2500 | 300 | 12.39 | 24 | No | $3.5 \times 10^3 / 7.1 \times 10^4$ |
| $BiBO_3$ | 3000 | 300 | 15.58 | 24 | No | $5.1 \times 10^3 / 7.3 \times 10^4$ |
| $Mn_2C$ | 2500 | 400 | 11.37 | 21 | No | $1.6 \times 10^4 / 5.5 \times 10^4$ |
| InCl | 500 | 75 | 33.98 | 20 | Yes | $1.8 \times 10^4$ |
| $VO_2$ | 1500 | 400 | 11.86 | 16 | No | $1.8 \times 10^4 / 8.8 \times 10^4$ |
| $Al_2O_3$ | 1500 | 400 | 10.05 | 16 | No | $7.3 \times 10^3 / 9.1 \times 10^4$ |
| $TiO_2$ | 2500 | 400 | 12.16 | 16 | No | $2.4 \times 10^3 / 6.4 \times 10^4$ |
| $Na_3N$ | 75 | 75 | 29.74 | 16 | Yes | $9.6 \times 10^3$ |
| GeTe | 1150 | 400 | 32.35 | 16 | No | $9.6 \times 10^3 / 3.3 \times 10^4$ |
| $SiO_2$ | 3000 | 300 | 18.15 | 24 | Yes | $1.1 \times 10^4$ |
| $BaTiO_3$ | 3000 | 300 | 15.59 | 30 | Yes | $4.8 \times 10^3$ |
| $CaCO_3$ | Range | 400 | Range | 30 | Yes | $1.6 \times 10^5$ ($n_{grid} = 20$) |
| $Fe_{80}B_{20}$ | 900 | 400 | 10.21 | 16 | No | $- / 3.9 \times 10^6$ ($n_{grid} = 20$) |
| BN | Range | 350 | Range | 8 | Yes | $3.1 \times 10^6$ |